\documentclass{tcibook}
\usepackage{fancyhea}
\usepackage{work}
\usepackage{bm}       
\usepackage{graphicx}
\usepackage{multirow}
\usepackage{lineno}
\usepackage{threeparttable}
\usepackage[normalem]{ulem}
\usepackage{color}
\usepackage[colorlinks = true,
            linkcolor = blue,
            urlcolor  = blue,
            citecolor = blue,
            anchorcolor = blue]{hyperref}
\usepackage[resetlabels]{multibib}
\newcites{ned}{Noble Element Detectors -- References}

\newcommand\summaryname{Abstract}
\newenvironment{Abstract}%
    {\small\begin{center}%
    \bfseries{\summaryname} \end{center}}

\def\beq{\begin{equation}}
\def\eeq#1{\label{#1}\end{equation}}
\def\eeqn{\end{equation}}

\newenvironment{Eqnarray}%
   {\arraycolsep 0.14em\begin{eqnarray}}{\end{eqnarray}}
\def\beqa{\begin{Eqnarray}}
\def\eeqa#1{\label{#1}\end{Eqnarray}}
\def\eeqan{\end{Eqnarray}}

\let\bar=\overbar

\def\lsim{\mathrel{\raise.3ex\hbox{$<$\kern-.75em\lower1ex\hbox{$\sim$}}}}
\def\gsim{\mathrel{\raise.3ex\hbox{$>$\kern-.75em\lower1ex\hbox{$\sim$}}}}

\def\del{\partial}
\def\Dslash{\not{\hbox{\kern-4pt $D$}}}
\def\dslash{\not{\hbox{\kern-2pt $\del$}}}
\def\pslash{\not{\hbox{\kern-2pt $p$}}}
\def\ETmiss{\not{\hbox{\kern-4pt $E$}}_T}

\def\Dlr{\mathrel{\raise1.5ex\hbox{$\leftrightarrow$\kern-1em\lower1.5ex\hbox{$D$}}}}

\def\MSB{{\bar{M \kern -2pt S}}}
\def\msb{{\bar{\scriptsize M \kern -1pt S}}}

\def\drb{{\bar{\scriptsize D \kern -1pt R}}}

\def\authorlist#1#2{
    \vskip 0.4in
\begin{center}\begin{large} {\bf  #1 } \end{large}
    \vskip 0.2in
              #2
     \vskip 0.2in
   \end{center}
}
\def\Address#1{\begin{center}{ \it #1} \end{center}}

\begin{document}


\pagenumbering{roman}

\parindent=0pt
\parskip=8pt
\setlength{\evensidemargin}{0pt}
\setlength{\oddsidemargin}{0pt}
\setlength{\marginparsep}{0.0in}
\setlength{\marginparwidth}{0.0in}
\marginparpush=0pt


\pagenumbering{arabic}

\renewcommand{\chapname}{chap:intro_}
\renewcommand{\chapterdir}{.}
\renewcommand{\arraystretch}{1.25}
\addtolength{\arraycolsep}{-3pt}

































\setcounter{chapter}{7} 


\chapter{Report of the Topical Group on Noble Element Detectors for Snowmass 2021}

\authorlist{C.~E.~Dahl$^{1,2}$, R.~Guenette$^{3,4}$, J.~L.~Raaf$^{2}$}
{D.~Akerib$^{5}$, J.~Asaadi$^{6}$, D.~Caratelli$^{7}$, E.~Church$^{8}$, M.~Del Tutto$^{2}$, A.~Fava$^{2}$, R.~Gaitskell$^{9}$, G.~K.~Giovanetti$^{10}$, G.~Giroux$^{11}$, D.~Gonzalez Diaz$^{12}$, E.~Gramellini$^{2}$, S.~Haselschwardt$^{13}$, C.~Jackson$^{8}$, B.~J.~P.~Jones$^{6}$, A.~Kopec$^{14}$, S.~Kravitz$^{13,15}$, H.~Lippincott$^{7}$, J.~Liu$^{16}$, C.~J.~Martoff$^{17}$, A.~Mastbaum$^{18}$, C.~Montanari$^{2}$, M.~Mooney$^{19}$, K.~Ni$^{20}$, L.~Pagani$^{21}$, O.~Palamara$^{2}$, L.~Pandola$^{22}$, R.~Patterson$^{23}$, S.~Pereverzev$^{24}$, X.~Qian$^{25}$, C.~Savarese$^{26}$, P.~Sorensen$^{13}$, C.~Stanford$^{3}$, A.~Szelc$^{27}$, M.~Szydagis$^{28}$, S.~Westerdale$^{29}$, J.~Xu$^{24}$, J.~Zennamo$^{2}$, J.~Zettlemoyer$^{2}$, C.~Zhang$^{25}$
}

\Address{
$^{1}${Northwestern University, Evanston, IL 60208, USA}\\
$^{2}${Fermi National Accelerator Laboratory, Batavia, IL 60510, USA}\\
$^{3}${Harvard University, Cambridge, MA 02138, USA} \\
$^{4}${University of Manchester, Manchester M13 9PL, UK}\\
$^{5}${SLAC National Accelerator Laboratory, Menlo Park, CA 94025, USA}\\
$^{6}${University of Texas at Arlington, Arlington, TX 76019, USA}\\
$^{7}${University of California Santa Barbara, Santa Barbara, CA 93106, USA}\\
$^{8}${Pacific Northwest National Laboratory, Richland, WA 99352, USA}\\
$^{9}${Brown University, Providence, RI 02912, USA}\\
$^{10}${Williams College, Williamstown, MA 01267, USA}\\
$^{11}${Queen's University, Kingston, Ontario K7L 3N6, Canada}\\
$^{12}${Universidade de Santiago de Compostela, IGFAE, Santiago de Compostela ES-15782, Spain}\\
$^{13}${Lawrence Berkeley National Laboratory, Berkeley, CA 94720, USA}\\
$^{14}${Purdue University, West Lafayette, IN 47907, USA}\\
$^{15}${University of Texas at Austin, Austin, TX 78712, USA}\\
$^{16}${University of South Dakota, Vermillion, SD 57069, USA}\\
$^{17}${Temple University, Philadelphia, PA 19122, USA}\\
$^{18}${Rutgers University, New Brunswick, NJ 08901, USA}\\
$^{19}${Colorado State University, Fort Collins, CO 80523, USA}\\
$^{20}${University of California San Diego, La Jolla, CA 92093, USA}\\
$^{21}${University of California Davis, Davis, CA 95616, USA}\\
$^{22}${Laboratori Nazionali del Sud, Istituto Nazionale di Fisica Nucleare, I-95123 Catania, Italy}\\
$^{23}${California Institute of Technology, Pasadena, CA 91125, USA}\\
$^{24}${Lawrence Livermore National Laboratory, Livermore, CA 94551, USA}\\
$^{25}${Brookhaven National Laboratory, Upton, NY 11973, USA}\\
$^{26}${Princeton University, Princeton, NJ 08544, USA}\\
$^{27}${University of Edinburgh, Edinburgh EH8 9YL, UK}\\
$^{28}${State University of New York at Albany, Albany, NY 12222, USA}\\
$^{29}${University of California at Riverside, Riverside, CA 92521, USA}\\
}

\begin{Abstract}
Particle detectors making use of noble elements in gaseous, liquid, or solid phases are prevalent in neutrino and dark matter experiments and are also used to a lesser extent in collider-based particle physics experiments. These experiments take advantage of both the very large, ultra-pure target volumes achievable and the multiple observable signal pathways possible in noble-element based particle detectors. As these experiments seek to increase their sensitivity, novel and improved technologies will be needed to enhance the precision of their measurements and to broaden the reach of their physics programs. The areas of R\&D in noble element instrumentation that have been identified by the HEP community in the Snowmass process are highlighted by five key messages: IF08-1) Enhance and combine existing modalities (scintillation and electron drift) to increase signal-to-noise and reconstruction fidelity; IF08-2) Develop new modalities for signal detection in noble elements, including methods based on ion drift, metastable fluids, solid-phase detectors and dissolved targets.  Collaborative and blue-sky R\&D should also be supported to enable advances in this area; IF08-3) Improve the understanding of detector microphysics and calibrate detector response in new signal regimes; IF08-4) Address challenges in scaling technologies, including material purification, background mitigation, large-area readout, and magnetization; and IF08-5) Train the next generation of researchers, using fast-turnaround instrumentation projects to provide the design-through-result training no longer possible in very-large-scale experiments. This topical group report identifies and documents recent developments and future needs for noble element detector technologies. In addition, we highlight the opportunity that this area of research provides for continued training of the next generation of scientists.Particle detectors making use of noble elements in gaseous, liquid, or solid phases are prevalent in neutrino and dark matter experiments and are also used to a lesser extent in collider-based particle physics experiments. These experiments take advantage of both the very large, ultra-pure target volumes achievable and the multiple observable signal pathways possible in noble-element based particle detectors. As these experiments seek to increase their sensitivity, novel and improved technologies will be needed to enhance the precision of their measurements and to broaden the reach of their physics programs. The areas of R\&D in noble element instrumentation that have been identified by the HEP community in the Snowmass whitepapers and Community Summer Study are highlighted by five key messages: IF08-1) Enhance and combine existing modalities (scintillation and electron drift) to increase signal-to-noise and reconstruction fidelity; IF08-2) Develop new modalities for signal detection in noble elements, including methods based on ion drift, metastable fluids, solid-phase detectors and dissolved targets.  Collaborative and blue-sky R\&D should also be supported to enable advances in this area; IF08-3) Improve the understanding of detector microphysics and calibrate detector response in new signal regimes; IF08-4) Address challenges in scaling technologies, including material purification, background mitigation, large-area readout, and magnetization; and IF08-5) Train the next generation of researchers, using fast-turnaround instrumentation projects to provide the design-through-result training no longer possible in very-large-scale experiments. This topical group report identifies and documents recent developments and future needs for noble element detector technologies. In addition, we highlight the opportunity that this area of research provides for continued training of the next generation of scientists.
\end{Abstract}

\newpage
\section{Executive Summary} 

\begin{figure}[t]
\begin{center}
\includegraphics[width=\textwidth]{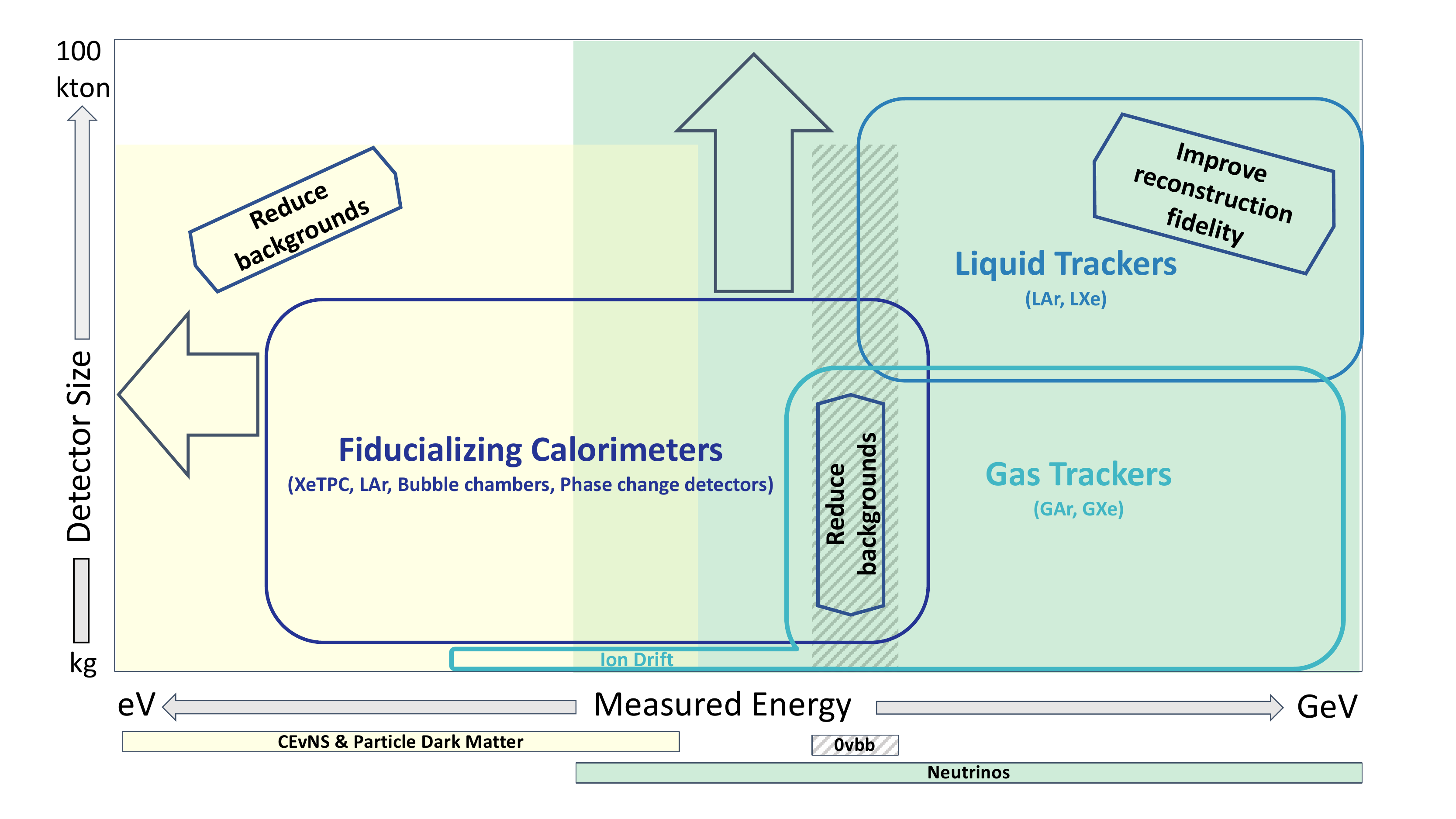}
\caption{Summary of the multiple physics goals (shaded regions) targeted by noble element detectors and the variety of detector technologies (boxes) used to meet those goals, with relevant energy ranges and detector sizes indicated.  The efforts described in this report aim to extend the reach (arrows) and improve the performance (angle boxes) of these detectors.}
\label{sum_fig}
\end{center}
\end{figure}

Particle detectors making use of noble elements in gaseous, liquid, or solid phases are prevalent in neutrino and dark matter experiments and are also used to a lesser extent in collider-based particle physics experiments. These experiments take advantage of both the very large, ultra-pure target volumes achievable and the multiple observable signal pathways possible in noble element particle detectors. Figure \ref{sum_fig} illustrates different goals for noble element detector technologies in function of targeted energies and detector size. As these experiments seek to increase their physics reach, novel and improved technologies will be needed to enhance the precision of their measurements and to broaden their physics programs. The priority research directions (PRDs) and thrusts identified in the 2019 Report of the Office of Science Workshop on Basic Research Needs for HEP Detector Research and Development (BRN report)~\citened{if08-osti_1659761} are still relevant in the context of this Snowmass 2021 topical group. The areas of R\&D in noble element instrumentation that have been identified by the HEP community in the Snowmass whitepapers and Community Summer Study align well with the BRN report PRDs, and are highlighted by five key messages (with IF-wide themes in bold):
\begin{itemize}
    \item[IF08-1:] {Enhance and combine existing modalities (light and charge) to {\bf increase signal-to-noise and reconstruction fidelity}.}
    \item[IF08-2:] {{\bf Develop new modalities for signal detection} in noble elements, including methods based on ion drift, metastable fluids, solid-phase detectors and dissolved targets.  Collaborative and blue-sky R\&D should also be supported to enable advances in this area.}
    \item[IF08-3:] {Improve the understanding of {\bf detector microphysics} and calibrate {\bf detector response in new signal regimes}.}
    \item[IF08-4:] {{\bf Address challenges in scaling technologies}, including material purification, background mitigation, large-area readout, and magnetization.}
    \item[IF08-5:] {{\bf Train the next generation of researchers}, using fast-turnaround instrumentation projects to provide the design-through-result training no longer possible in very-large-scale experiments.}
\end{itemize}

This topical group report identifies and documents recent developments and future needs for noble element detector technologies. In addition, we highlight the opportunities that this area of research provides for continued training of the next generation of scientists.

\newpage

\section{Enhancing existing modalities}
Noble element detectors developed for neutrino physics and dark matter searches record mainly the charge from the ionization electrons and the light from the scintillation produced by the passage of charged particles through the medium. The technologies to read out the charge in neutrino experiments have mostly been based on wire readouts, such as ICARUS~\citened{ICARUS:2004wqc, ICARUS-T600:2020ajz}, ArgoNeuT~\citened{Anderson:2012vc}, MicroBooNE~\citened{MicroBooNE:2016pwy, MicroBooNE:2016ptd}, LArIAT~\citened{LArIAT:2019kzd}, SBND~\citened{SBND:2020scp}, DUNE first far detector module~\citened{DUNE:2020txw}, and EXO~\citened{Auger:2012gs}, while charge measurements in dark matter searches rely on gain mechanisms such as gas electroluminescence~\citened{Aalbers:2022dzr,DarkSide-20k:2017zyg} and proportional gain~\citened{NEWS-G:2022kon}. To read out scintillation light, both neutrino and dark matter experiments have focused on the use of PMTs and SiPMs with coverage ranging from sub-percent to up to 50$\%$ levels.

It is clear that for the future, advances in charge and light detection capabilities are highly desirable, and to this end a range of new approaches have been proposed, discussed in more detail below. 

\subsection{Pixels} 
Although the concept of wire-based readout has been proven and has had wide usage in neutrino detectors \citened{ICARUS:2004wqc,MicroBooNE:2016pwy,DUNE:2020lwj}, it has an intrinsic limitation in resolving ambiguities, resulting in potential failures of event reconstruction. In addition, the construction and mounting of massive anode plane assemblies to host thousands of finely spaced wires poses significant and costly engineering challenges. For these reasons, a non-projective readout presents many advantages, but the large number of readout channels and the low-power consumption requirements have posed conisderable challenges for applicability in liquid noble TPCs. The number of pixels compared to the number of corresponding sense wires will be two or three orders of magnitude higher for equal spatial resolution, with an analogous increase of the number of signal channels, data rates, and power dissipation. The endeavour to build a low-power pixel-based charge readout for use in LArTPCs has independently inspired the LArPix~\citened{Dwyer:2018phu} and Q-Pix~\citened{Nygren:2018rbl} consortia to pursue complimentary approaches to solving this problem.

There a several benefits of a native 3D readout for noble element TPCs. An intrinsic 3D readout offers an increased event reconstruction efficiency and purity, and the ability to accurately reconstruct final state topologies with greater detail, as shown in Ref.~\citened{Adams:2019uqx}. It was also demonstrated that a pixel-based readout has enhanced capabilities to reconstruct low-energy neutrino events ($\mathcal{O}(\leq 5)$~MeV) with improvements in overall data rates and signal fidelity~\citened{supernova_paper} . Additional work is ongoing to fully explore the physics potential realized by a pixel-based noble element readout, but initial studies are promising.

The technical requirements for a pixel readout are broadly classified in terms of the readout noise, power, and reliability in a cryogenic environment. While the specifics are defined by their bespoke application, the general themes of power requirements include: $<10$~W/m$^2$ average power with mm-scale pixels, $\leq500$ e$^-$ equivalent noise charge (ENC), and sub-\% failure tolerance across a system with millions to billions of channels. Along with these requirements on the pixel electronics comes some on system reliability and scalability. Robust input/output (I/O) architectures are needed to faithfully bring the data from the large number of channels created by a pixel readout to a central data logger. While this challenge is not unique to pixel readouts, there is a particular demand on the readout architecture to not introduce any noise, since a low threshold is needed to achieve the sensitivity gains that pixels aim for (see below). To scale these devices up to larger experiments, groups must leverage commercial methods for mass production. This includes targeting ASIC and printed circuit board (PCB) manufacturing processes that are well-suited for low cost and high reliability. Furthermore, in order to detect scintillation light, chambers must be equipped with a light collection system sensitive to VUV light  (128-175~nm), ideally to be integrated with the pixel plane~\citened{Barman:2021scx, solar}.

As the pixel technology continues to mature, there is a drive to push the detection threshold limit to lower values. Future neutrino experiments aim to have enhanced sensitivity to supernova and solar neutrino events, which will require energy thresholds around, and potentially below, 1 MeV. Such lower thresholds could also allow for the application of pixel-based noble element TPCs to explore beyond-the-standard-model (BSM) physics, as well as for probing areas of low-threshold detection (e.g., dark matter~\citened{arxiv.2203.08821} and neutrinoless double beta decay~\citened{arxiv.2203.14700}) which have thus far been focused only on the use of the secondary scintillation light for charge readout. As the thresholds are lowered, new challenges arise in increasing the ``dynamic range'' of circuits and in ensuring that data rates stay manageable. The need to find and share common solutions within the community of researchers pursuing pixel-based readout for noble element TPCs is a key instrumentation challenge. Often arbitrary barriers due to various intellectual property concerns of the ASIC foundries cause multi-institutional collaborations to be difficult if not impossible, which slows the progress of R\&D. Platforms such as the CERN R\&D collaboration have found ways to overcome this and have been essential for delivering technologies used by the current generation of large high-energy physics experiments. The creation of a similar platform within the US would allow for the best ideas to come together into a final viable design with efficient use of available resources. This structure would greatly enhance cooperative technology development across multiple experiments.

\subsection{Light collection}
The information carried by photons is critical for a wide range of physics measurements in noble element–based detectors, providing a crucial means by which to perform detector triggering as well as position and energy reconstruction and identifying interactions of interest. It is essential to fully leverage this information in next-generation measurements, including neutrino interactions from low-energy coherent scattering (CEvNS) up to the GeV energy scale, neutrino astrophysics, and Beyond the Standard Model physics searches such as for low-mass dark matter and neutrinoless double beta decay. These efforts will require substantial (even as high as 100-fold) increases in light collection, to enable percent or sub-percent level energy resolution, mm-scale position resolution, low-energy detector readout triggering, and/or highly efficient particle identification, including for events around  $\mathcal{O}(\leq 1)$~MeV (keV) energies in detectors at the 10~kton (100~ton) scale. In the context of the broader program of noble element detectors, enhancements in photon collection will lead to dramatic improvements in event reconstruction precision and particle identification, in a broader range of physics signatures afforded by lower trigger thresholds, and in precision timing to unlock new handles for beam-related events.

Measurement of the light signals, however, presents a major challenge with currently available technologies. For example, in large-scale liquid argon neutrino detectors, it is typical to collect $<$1$\%$ of the produced photons. This limitation is driven in large detectors by geometric considerations and other active components, total heat load, data volume, and the cost of instrumented surface area. The efficiency of the photodetectors and of the wavelength shifters used to convert VUV scintillation light to optical wavelengths for detection also play a role, as do the shortcomings of currently available devices such as noise, dark rate, and after-pulsing. In consideration of the very large scales of next-generation experiments — in both run time and target mass — significant R$\&$D is needed to move beyond these limitations and enable the future physics program. While the overall needs are common across a wide range of physics goals, specific measurements will likely require a case-specific optimization of overall photon statistics, timing, and pulse shape discrimination performance. This demands a broad effort to develop a comprehensive and robust simulation of optical photon production, transport, and detection, including characterization of the optical properties of detector materials.

Several promising approaches to improving light collection can address one or more of the limitations noted above, and taken together, provide a set of complementary tools for next-generation experiments: photon collection efficiency may be improved by imaging a large volume on a small active detector surface using novel lensing technologies~\citened{Villalpando_2020}, deployment of reflective and wavelength-shifting passive surfaces \citened{Boulay:2021njr,Abraham:2021otn,Kuzniak:2020oka}, bulk wavelength-shifting through dissolved dopants \citened{Akimov:2017ble,Wahl:2014vma,Neumeier:2015ori,Peiffer:2008zz,Segreto:2020qks}, improvement of photon detectors \citened{Villalpando:2020tsc} and photon transport efficiency, or the conversion of photons into ionization charge for readout using a TPC \citened{ANDERSON1986361}. These strategies couple to many other elements of detector design and physics performance, including photon detector technologies, low-energy/low-background physics, simulation tools, and readout and data acquisition R$\&$D. These approaches, implemented individually or in combination as optimized using a detailed microphysical simulation, will afford radical enhancements in the capabilities of future detectors, especially in the challenging low-energy regime near and below 1~MeV. Continuing to develop detailed and accurate light propagation simulation tools (e.g.~\citened{Szydagis_2021}) will be essential to assist the discussed R$\&$D.

\subsection{Extreme low thresholds (electron counting)} 
The lowest energy phenomena that can be studied with existing noble-element modalities, including low-mass dark matter, reactor neutrinos, and natural (e.g. solar) neutrinos, require detectors that are sensitive to single ionization electrons.  Such detectors are sensitive to $\mathcal{O}$(10~eV) electronic recoils and $\mathcal{O}$(100~eV) nuclear recoils, but lack the scintillation-dependent nuclear and electronic recoil discrimination present at higher energies. Two-phase argon or xenon detectors, which achieve this sensitivity through gas-phase electroluminescence, are well developed for heavy WIMP searches, but their signal production mechanisms and backgrounds below $\mathcal{O}$(keV) need further investigation. Liquid neon deserves renewed investigation to complement gas-phase efforts taking advantage of its intrinsic radiopurity and favorable kinematics for recoil energy transfer from light dark matter and low-energy neutrinos~\citened{NEWS-G:2021mhf}. A new class of compact $\mathcal{O}$(100~kg) low-threshold (sub-keV) noble element detectors will offer complementary physics opportunities to large (100~tonne) noble liquid detectors in dark matter and neutrino physics, while being competitive with other low-threshold detector technologies that are more difficult to scale up in target mass. 

Without nuclear and electronic recoil discrimination, systematic backgrounds and radioactivity obscure typically background-free nuclear recoil event searches, and the radiopurity of detector materials, particularly photosensors, becomes critical. Better liquid or gas purification techniques (e.g. cryogenic distillation) drastically reduce beta and gamma backgrounds stemming from $^{3}$H, $^{39}$Ar, $^{85}$Kr, and the $^{220,222}$Rn decay chains. Cosmogenic activation rates must be further studied and considered for handling detector materials above ground. Beyond background particle interactions, high rates of single- and few-electron signals are observed. Such spurious electrons have defied clear explanation and appear related to charge build-up on surfaces or in unknown chemical interactions, among other potential effects \citened{DarkSide:2018bpj,RED-100:2019rpf,LUX:2020vbj,Kopec:2021ccm,XENON:2021myl,Pereverzev:2021zoi}. Dedicated R$\&$D is needed to better understand the sources of these backgrounds and to develop mitigation techniques. Fast and efficient gas purification, liquid purification technologies, cleaner alternative detector materials, and various electric field configurations must be explored to optimize signal measurement efficiency and reduce backgrounds. Electroluminescence in a single-phase noble element detector, either high-pressure gas \citened{10.1007/3-540-26373-X_22} or liquid \citened{Wei:2021nuk, Kuger:2021sxn} , offers a thermodynamically simpler possibility for single-electron detection, without the hypothesized electron-trapping at the liquid-gas interface in two-phase LXe detectors. 

The cross-cutting challenges in Sec.~\ref{IF08-xcut}, including in- and ex-situ calibrations and development of doping schemes, are particularly relevant to single-electron-sensitive experiments.  Validation of sensitivity below $\mathcal{O}$(keV), including measurements of new phenomena such as the Migdal effect \citened{Ibe:2017yqa,Bell:2021ihi}, will require substantial effort developing new calibration sources and techniques.  Doping, with noble or non-noble dopants, can both improve operation and increase the physics reach of single-electron-sensitive detectors by boosting ionization yield.

\subsection{Charge gain}
Lower detection thresholds in track-reconstructing detectors (where the electroluminescence techniques of the previous section are less useful) may be achievable through amplification of the ionization signal in the form of charge gain, typically achieved in the gaseous phase of argon and xenon detectors.
Gas-phase charge gain without scintillation is well established, and used by the NEWS-G collaboration \citened{Giomataris:2008ap,Gerbier:2014jwa} to search for low-mass WIMP-like particles using spherical proportional counters (SPCs) filled with gases such as neon, methane, and helium.
Multiple innovative methods are being developed to either enhance the capabilities of charge amplification in gaseous detectors (e.g. to achieve stable charge gain while retaining the primary scintillation channel) or to enable amplification directly in the liquid phase. Active R$\&$D efforts on this front are described below.

{\bf Electron multiplication in liquid argon TPC detectors:} Enabling charge amplification directly in liquid argon would expand the physics reach of liquid argon detectors, reducing thresholds to $<$ 100 keV in energy and opening up new areas of research in processes such as dark matter and CE$\nu$NS searches. Achieving charge amplification in liquid is significantly more challenging than in gas due to the denser medium and higher electric field thus required. Past work on this idea~\citened{Titov:2013hmq,Bondar:2005wx,Bressi:1991yj,Kim:2002wv}, while promising, has not yet reached a level of maturity necessary to enable the technological advances needed for physics measurements. Benefits of direct amplification in liquid are a potentially improved detector stability (due to the lack of a gas-liquid surface) and detector scalability. Active R$\&$D through the LArCADE program is exploring this possibility through the implementation of strong local electric fields with tip-arrays instrumented at the TPC’s anode readout. 

{\bf Scintillating and quenched gas mixtures for high-pressure gaseous TPCs:} While there is a rich history of R\&D in gaseous detector readout electronics, much remains to be understood and optimized at the high pressures and large scales sought by experiments. The realization of a stable, VUV-quenched gain, scintillation-compatible, 10-15~bar TPC remains elusive. Two main approaches are pursued to enable these objectives, distinguished by their scintillation wavelength range: infra-red and near-UV or visible readout.  An ongoing program of R$\&$D 
aims to systematically map the space of scintillating gas mixtures of argon with admixtures of xenon, nitrogen, hydrocarbons, and fluorinated compounds. The potential of new `ad hoc' mixtures for pure and low-quenched gases also impacts the development of new ideas, for instance in DUNE's ND-GAr detector~\citened{DUNE:2021tad, DUNE:2022yni} where Ar-CF$_{4}$ is being considered as a scintillating gas for providing the start time (T$_{0}$). In such a case, the screening of the secondary scintillation impinging on the photosensor plane might be critical, something that a GEM (conveniently optimized) can provide. Further, independent measurements and calculations also suggest that stable scintillation on very thick structures (5-10~mm thick) is likely to be possible in liquid phase \citened{Buzulutskov:2018vgg,NEXT:2022ijl,Amedo:2021rsv,Borisova:2021bio}.

\section{New modalities} 
As a detection medium, noble elements present unique opportunities beyond the collection of scintillation photons and ionized electrons.  The modalities described in this section find new ways to utilize the monolithic, ultra-pure elemental detection medium provided by noble elements, extending the reach of noble-element-based detectors to new signal regimes and enabling new methods of background discrimination in rare event searches.

\subsection{Ion Detection and Micron-scale Track Reconstruction}
The ability to reconstruct ionization tracks at micron- and sub-micron spatial resolution is the key to many currently unsolved detector challenges, including directional dark matter detection ($\mathcal{O}$($10^{-6}$g/cm$^2$) spatial resolution required), discrimination of single-electron backgrounds in 0$\nu\beta\beta$ searches ($\mathcal{O}$($10^{-1}$g/cm$^{2}$) spatial resolution required), and potentially for detection of supernova and solar neutrino events in very large-scale neutrino detectors ($\mathcal{O}$(sub-mm) spatial resolution required).  Attempts at direct (TPC-style) high-resolution reconstruction universally rely on ion drift rather than electron drift to escape the resolution-limiting effects of electron diffusion over large drift distance.  The drifting ions may be either positive ions of the target itself~\citened{Amaro:2022gub}, or a positively or negatively ionized dopant~\citened{Marques:2022wyr, Martoff:2004wf,Martoff:2000wi}.  Imaging the ion arrival on the cathode (or anode for negative ions) plane can also take many forms, including CCDs in gas phase detectors~\citened{Amaro:2022gub} and long-time-scale fluorescence activation by ions in liquid phase detectors~\citened{Jones:2022moh}.  Selective readout of the imaging plane is often a necessary component of the large-scale application of these techniques, due to both pileup and data throughput limitations.  Selective readout may be directed by real-time ``low-resolution'' electron-drift-based imaging~\citened{Jones:2022moh}.

It may also be possible to sense ion tracks indirectly via their interaction with drifting electrons, and the corresponding impact on standard TPC observables.  Columnar recombination models predict variation in relative ionization and scintillation yields based on the orientation a track with respect to the applied drift field.  Several efforts are investigating the magnitude of this effect and how it may be applied for directional dark matter detection~\citened{Agnes:2021zyq,Wojcik:2015zqp}, as well as its impact in high energy neutrino experiments~\citened{ArgoNeuT:2013kpa}.

Ion transport and detection is also a key consideration for barium tagging, a more direct approach to 0$\nu\beta\beta$ background discrimination that seeks to identify the barium ions left behind by the double-beta decay of $^{136}$Xe~\citened{Mong:2014iya,Craycraft:2019hid,Twelker:2014zsa,Flatt:2007aa,Murray:2019snw,NYGREN20162,Jones:2016qiq,Thapa:2019zjk,thapa2021demonstration,Rivilla:2020cvm,Herrero-Gomez:2022uci,NEXT:2021idl,Bainglass:2018odn,Brunner:2014sfa,nEXO:2018nxx,McDonald:2017izm}.  
There are several key instrumentation requirements for a workable barium tagging technology. First, the system must collect and detect barium ions or atoms with high efficiency and selectivity, since a significant level of inefficiency amounts to wasted exposure time of active isotope.  The sensor must be uninhibited by spurious signals from ambient background atoms or ions.  A detection limit of exactly one barium ion must be reached by the senor of choice, in the environment of the sensing region. The full fiducial volume of the detector must be accessible by the barium tagging system.  A spatio-temporal coincidence between the ion collected and the electrons emitted in the event must be maintained. And, the barium tagging system must be realized in such a way that it does not introduce radio-impurity or compromise other key detector functions such as energy resolution.
%
Detection of single Ba atoms and ions has now been demonstrated using several techniques, but continued R\&D is needed to achieve and quantify the high efficiency and selectivity needed for a practical barium tagging application.  Development of methods to transport  / extract barium from ions or atoms in either LXe or GXe is a critical step, with ongoing work focused on radiofrequency carpets and funnels, actuated cryoprobe insertions, wide-area laser scanned cathodes, and ion mobile surfaces.


\subsection{Metastable fluids}
Metastable fluid detectors amplify the energy deposited in particle interactions with the stored free energy in a superheated or supercooled liquid target.  That amplification can be made selective by matching the different energy-loss mechanisms and length scales for signal and background interactions with the relevant phase-change thermodynamics, for example allowing bubble chambers to detect $\mathcal{O}$(1 keV) nuclear recoils (e.g. from dark matter or coherent neutrino scattering) while being completely blind to electron recoil backgrounds.  Instrumentation efforts in this area typically focus on (1) extending phase-change based discrimination to new signal regimes, and (2) improving control of spurious phase-change nucleation to enable larger quasi-background-free exposures.

While metastable fluid detectors are not restricted to noble elements, noble-liquid bubble chambers present unique opportunities.  The addition of scintillation detection to a bubble chamber is a powerful tool for discriminating against high-energy bubble-nucleating backgrounds \citened{Baxter:2017ozv}, and at the same time the limited energy-loss pathways available in a noble liquid target result in orders-of-magnitude improvements in low-energy (sub-keV) electron recoil discrimination \citened{Giampa:2021wte}.  The SBC Collaboration is actively developing the liquid-noble bubble chamber technique~\citened{Alfonso-Pita:2022akn}, with focus on three bubble chamber firsts: cryogenic operation of a large ``clean'' bubble chamber, stable superheating at $\mathcal{O}$(100 eV) thresholds, and precision nuclear recoil calibrations with $\mathcal{O}$(10 eV) resolution.  The last of these will involve both the development of new low-energy nuclear recoil calibration schemes, such as Thomson scattering by high-energy gammas and nuclear recoils from gamma emission following thermal neutron capture, and the development of the analysis techniques needed to combine diverse calibration data to constrain nucleation thresholds at the required resolution.


Freon-filled bubble chambers, such as those operated by the PICO Collaboration \citened{PICO:2019vsc}, continue to play a key role in high-mass dark matter detection, enabling quasi-background-free nuclear recoil detection in targets with high spin-dependent and low spin-independent cross-sections.  This allows the exploration of orders-of-magnitude more dark matter parameter space before reaching the “neutrino fog” than can be achieved in noble liquid targets, with strong physics motivation for freon bubble chambers out to kiloton-year exposures and beyond~\citened{Akerib:2022ort}. Those exposures cannot be achieved without the development of new bubble chamber designs that are more scalable than the current fused-silica chambers, while maintaining (or improving on) current chambers’ low spurious nucleation rate.  This requires studies of bubble nucleation on surfaces and new bubble-imaging methods (e.g. acoustic imaging), both of which directly benefit liquid-noble bubble chambers as well.  Larger exposures will also require the development of active neutron vetos compatible with the bubble chamber environment.

A third application of metastable fluids is the detection of proton recoils in water, providing a light target with nearly pure spin-dependent coupling that is kinematically matched to low-mass dark matter.  Water is a notoriously difficult fluid to use as a bubble chamber target, but Snowball chambers \citened{Szydagis:2018wjp} sidestep this roadblock by supercooling the target rather than superheating it.  An entirely new particle detection technology, Snowball chambers face the same instrumentation challenges as SBC and PICO:  surface nucleation must be mitigated, and both the threshold and discrimination power of the technique must be calibrated.

Finally, there is renewed interest in accelerator-based bubble chamber experiments to measure neutrino cross sections on light nuclei~\citened{Alvarez-Ruso:2022exy}.  While the requirements for these devices, including high delta-ray sensitivity and fast ($>$10-Hz) cycling, push in a different direction than the dark matter and CEvNS-motivated chambers, practical concerns including control algorithms, photography, and image analysis remain common between these efforts.

\subsection{New modalities in existing noble-element detectors}
It is highly desirable to find novel ways to take advantage of both the field-wide expertise that has developed around noble-element-based detectors and the world-class infrastructure surrounding existing and near-future searches for new physics. In general, after an experiment has achieved its scientific goals, the experimental community ideally will continue to leverage the infrastructure for future experiments. The simplest case of this general principle is perhaps that of upgrading an existing experiment. Previous examples include KamLAND $=>$ KamLAND-Zen, and the Darkside installation in the former Borexino CTF (Counting Test Facility).  There are many liquid noble installations around the world that can lend themselves to this sort of upgrade. As one example, the DEAP-3600 experiment is currently undergoing hardware upgrades for improved background rejection with future potential uses of the experimental infrastructure after the science run including a sensitive assay of $^{42}$Ar in underground argon or measurements of solar neutrinos. 

Currently, new ideas for upgrades of the LZ detector \citened{LZ:2019sgr}, after it completes its scientific goals in $\sim$2027, are under development.  The specific proposals are called HydroX and CrystaLiZe, both of which would benefit from leveraging the low-background installation with water tank shielding and liquid scintillator veto detectors, as well as the associated infrastructure. However, both would also require significant upgrades or modifications to the LZ inner detector. 

HydroX - The idea behind HydroX is to dissolve a hydrogen target in LZ to enable searches for very light dark matter.  Hydrogen is the ideal target for low-mass dark matter because it has the lowest atomic number of any element, and because its unpaired proton (and neutron in the case of deuterium) provides sensitivity to spin dependent couplings in the low mass range. As an upgrade for LZ, HydroX would leverage both the existing TPC (using xenon ionization and scintillation to detect proton recoils in LZ) and the major investment in the low background construction and radio-clean environment. Significant R$\&$D is still required to demonstrate the viability of this idea, primarily measuring detector properties of H$_2$-doped liquid xenon, including the signal yields of proton, electron, and xenon recoils, and understanding the cryogenics of H-doped LXe. A HydroX-like upgrade could also be envisioned for next generation dark matter efforts. 

CrystaLiZe - this is a proposal to crystallize the liquid xenon target of the LZ instrument. R$\&$D is underway to demonstrate the feasibility of this plan. If implemented, this upgrade could enable full tagging of radon-chain beta decay backgrounds, enabling crystaLiZe to be a neutrino-limited (rather than radon-limited) dark matter search. As with HydroX, this path forward would leverage the LZ infrastructure after LZ completes its science goals. A fundamental premise of this proposal is that crystalline xenon will have “the same” TPC-style particle detection capability as liquid xenon. Preliminary work shows the scintillation yields are identical. Next steps intend to confirm that the incident particle type discrimination is also possible.

A key point is that HydroX and CrystaLiZe appear to be fundamentally compatible with each other, that is, one could imagine doping a light element into a crystalline xenon target.

\section{Challenges in scaling technologies}
Next-generation large-scale detectors are planned to search for dark matter and 0$\nu\beta\beta$ and to study neutrinos from both artificial and natural sources. Achieving these goals generally requires (i) scaled-up target procurement and radiopurity and purification capabilities; (ii) large area photosensor development with low noise; 
(iii) high voltage and electric field capabilities compatible with multi-meter drifts; and (iv) studying the effects and techniques for operating large doped noble liquid/gas detectors. The discovery capabilities of these detectors could be extended further by coupling them with a magnetic field, such to enable charge discrimination and improve momentum measurement.

Concerning target procurement, argon detectors will need new sources of underground argon (UAr) to fill large LArTPCs (e.g. a DUNE low-background module \citened{arxiv.2203.08821} would need 7–17~kt). Reduction of $^{39}$Ar is expected to reach activities 1400 times lower than atmospheric argon (AAr), at a cost $\sim$3$\times$ that of AAr. Reduction in levels of $^{42}$Ar is expected to be orders of magnitude beyond the $^{39}$Ar reduction. Xenon detectors will need $\sim$100~t (and potentially up to kt-scales in the future) from commercial sources, representing $\sim$2 years of total annual output worldwide at costs which must be coordinated with vendors. Natural Xe has sufficiently low background for future searches, and a large target mass can be sold back, substantially reducing its cost below the upfront acquisition cost of $\$$1M/t. Isotopic separation can benefit rare event searches by separating out $^{136}$Xe from the target (for 0$\nu\beta\beta$-decay) and odd-neutron isotopes ($^{129,131}$Xe) reduced in $^{136}$Xe for a DM search. However this will significantly impact the cost of the Xe.

Backgrounds generally need to be reduced to the $\mathcal{O}$(1) event/exposure in the $<$200 keVnr energy range for dark matter searches and in the $\mathcal{O}$(MeV) range for neutrino experiments. This goal requires further radiopure detector development, including the identification of radiopure pressure vessels, cryostats, cables, connectors and photosensor materials. Significant progress has been made in low-background SiPM development, though Xe-sensitive SiPM systems (or Xe-compatible wavelength shifters) need to be improved (especially dark rates and effective QE of large area arrays) in order for Xe detectors to transition to PMTs to SiPMs. It is also necessary to reduce radioactive impurities in the target, and enrichment is needed for Xe-based 0$\nu\beta\beta$ searches. Cryogenic distillation has come a long way in this regard, and UAr has been shown to have substantially lower $^{39}$Ar contamination, and $^{42}$Ar may be negligible. Larger sources of UAr will be needed for applications much larger than Argo (300~t)~\citened{Back:2022maq}, and upgrades may be needed to Aria~\citened{DarkSide-20k:2021nia} to achieve a high throughput for similarly large volumes.  Cosmogenic activation of radioisotopes is also a challenge; new measurements may be needed to improve activation calculations.  Improved understanding of small isolated charge and light signals, whether originating from particle interactions, chemical interactions, or electrode surfaces, is also needed in order to address accidental-coincidence backgrounds in large TPCs. 

Large-area photon and charge detection techniques and their associated readouts are also needed. For light detection, this includes (i)  photosensor development with expanded light collection area and large-area wavelength shifters, (ii) development and production of low-background, low-noise cryogenic SiPMs, 
(iii) development of power-over fiber technology and low-power, high-multiplexing cold readout electronics for photodetection empowering high timing resolution, needed to achieve 4$\pi$ light detection with high surface-coverage for 4D tracking and dual calorimetry in a LArTPC, improving the PID and energy resolution. For charge detection, the development of large-area and low-noise electron multipliers is important to detect small signals in large detectors.

Combining this lower detection threshold with a magnetic field in the range of 0.5 to 1 Tesla in the fourth DUNE module would allow for an effective measurement of momentum, charge discrimination, better energy resolution for hadron showers, improved particle identification and identification of the starting point of low energy electrons. This has the potential to add significance to the physics output of the overall observatory, for instance by improving the sensitivity to CP violation with atmospheric neutrinos by 50$\%$~\citened{magnetized_physics}. Since external conventional magnets are not suited for large volume cryogenic detectors, a robust R$\&$D program is needed to evaluate alternatives based on superconducting magnets: these could range from warm superconductors requiring dedicated cryogenic infrastructure, such as MgB$_{2}$ to be operated at 15-20~K, to hot superconductors more directly integrable in the nitrogen cooling plant or directly in the liquid argon volume, such as YBCO that can be operated at the liquid nitrogen temperature of 77~K.
LArTPC performances in the presence of a magnetic field are currently being studied with ArCS (Argon detector with Charge Separation)~\citened{arcs}, an R$\&$D effort where a prototype LArTPC detector will be magnetized and placed on the Fermilab test beam to determine minimum field requirements to achieve particle charge separation and study electron diffusion in the presence of the field.

Larger noble element TPCs require higher high voltages (HV). New HV feedthrough (FT) designs are needed for these larger areas. Successful R$\&$D implementing a conventional HV FT was developed for the 4D-LArTPC DUNE module to obtain a homogeneous, vertical electric field of 500~V/cm over a 6.5~m drift with a 3x3m$^{2}$ anode plate~\citened{}{\textcolor{red}{(Missing ref)}}. Examples like this can be taken as a starting point to test and develop a new technology. A FT from a co-extruded multi-layer cable made of a single plastic material with an additional semi-resistive plastic layer between the insulation and ground can robustly and compactly deliver $>$100kV and generate electric fields within a detector. Such a cable can be manufactured by developing a semi-resistive plastic with tunable resistivity (between 107-1013~Ohm~cm for a thickness of 0.3 to 1~mm). A more quantitative understanding of HV breakdown thresholds in pure liquid noble elements is also desired. This should include the effect of surface preparation such as passivation or electropolishing.

While current dark matter and neutrino experiments have focused on pure, noble liquid targets (e.g argon and xenon), there is a significant interest in exploring the effects of doping liquid argon with xenon and other elements \citened{Fields:2020wge,Doke:1976zz,Akimov:2019eae,Peiffer:2008zz,Galbiati:2020eup,Vogl:2021rba,ICHINOSE1990354,MASUDA1989560,PhysRevB.54.15724,Cennini:1995ve,Suzuki:1986zd,Neumeier:2015ori}. These dopants are typically chosen for ease of light detection and increasing scintillation and ionization yields. At higher concentrations, dopants can also be favorable targets in their own right. For example, $^{136}$Xe doping in high quantities could allow for search of neutrinoless double beta decay \citened{arxiv.2203.14700}, hydrogen or hydrogenous compounds doped in liquid xenon provide a light nucleus with more efficient kinematic coupling to light dark matter, and nuclei with an odd number of nucleons can add spin-dependent sensitivity. Further research is needed to develop the capacity for stable, large-scale, high-purity doping and to measure the effects on signal production and propagation.


\section{\label{IF08-xcut}Cross-Cutting Challenges}

In order to be sensitive to a wide range of physics phenomena, we must be able to make accurate and precise measurements of charge, light, and/or heat, from interactions of interest within our detectors, which requires in turn, a good understanding of the inherent noise levels, calibrations, and microphysics associated with these gaseous and liquid noble detectors. The challenges therein are cross-cutting, touching many different areas of experimental physics. But this also means that we can take a wider view to leverage facilities that will benefit many experiments across multiple frontiers. This deep understanding of the detection characteristics and calibrations will be essential components of preparing next-generation gaseous/liquid noble detectors for cutting-edge physics measurements in both the Neutrino Frontier (NF) and Cosmic Frontier (CF). Relevant searches/measurements include dark matter searches, searches for neutrinoless double beta decay, coherent elastic neutrino-nucleus scattering measurements, probing neutrino oscillations for measurements of leptonic CP violation and other PMNS matrix parameters, measurements of supernova/solar neutrinos, and searches for proton decay and other forms of baryon
number violation.


\subsection{In-situ calibrations}
Looking toward the next generation of HEP experiments, there are a variety of requirements for instrumentation and calibration methodology in ensuring accurate and precise measurements of
charge and light in detectors making use of noble elements, such as argon or xenon. First, it should be noted that both electron recoils and sub-keV nuclear recoils in xenon and argon are of interest, as the full range of relevant experiments collectively probe both types of recoils. Lower detector energy thresholds, both in the bulk liquid and at the liquid-gas interface for two-phase (liquid target with gas phase for signal gain) technology, are needed to pursue the physics measurements described above. Establishing measurements of noble element properties (e.g., diffusion and electron-ion recombination) to sufficient levels prior to running large, next-generation noble element detectors is necessary, given that these experiments may not be able to make these measurements in situ. This includes addressing effects related to self-organized criticality and other dynamic effects at low energies arising from the interplay of condensed matter and chemical interactions in noble liquid detectors, such as accumulations/releases of excitation energy and Wigner crystallization, which are potential
backgrounds in rare event searches. 

Many challenges exist in pursuing the precise calibration of charge and light measurements in
next-generation noble element experiments. Greater background reduction at lower recoil energies is a significant challenge. Increasing light collection and quantum efficiencies well beyond current levels, in order to achieve lower energy thresholds and improve energy resolution, is a difficult problem. A variety of improvements are needed to increase light and charge collection efficiencies in liquid argon/xenon, including improving impurity modeling, purification methods, mitigation and accounting for material degassing, and estimating electron attachment rates for impurities. There are also currently significant uncertainties concerning how non-linear detector response becomes at the lowest recoil energies relevant to low-mass dark matter searches and coherent neutrino observations. While the development of atom-level simulations of charge and light yields (such as those being pursued by NEST~\citened{Szydagis_2021}) to improve modeling for noble element detectors will help address this challenge, better particle and detector models are needed to extract more information from data. Additionally, training the next generation of physicists to become experts in noble detector characterization/microphysics requires funding agencies to support continued work on detector calibrations as a foundational part of physics research; this will further develop the workforce necessary to enable the physics measurements of interest at relevant experiments. Finally, given the connections between condensed matter and nuclear physics effects and their manifestation in HEP detectors, it is important to improve the communication between the BES, NP, and HEP communities on these cross-cutting topics, which is currently lacking.

\subsection{Ex-situ detector characterization, facilities}
Flexible user facilities (not tied to any particular group nor experiment) play a key role in noble element R\&D, both for calibration needs where in situ measurements are insufficient and for short-term instrumentation tests where a permanent dedicated test stand is unnecessary. These facilities minimize duplication of efforts by providing community-wide resources to benefit multiple research efforts. Existing successful examples include the Test Beam Facility~\citened{LArIAT:2019kzd} and Liquid Noble Test Facility ~\citened{FNAl_larf} at Fermilab, and the Liquid Noble Test Facility (LNTF) at the IR2 experimental hall at SLAC~\citened{SLAC_lntf}.  

The Fermilab facilities offer existing fast-turnaround cryostats with both charge and light readout, enabling both sensor development and measurements of noble element properties with radioactive sources or the Fermilab test beam. The LNTF at SLAC allows users to bring their own cryostats, managing a number of common-use systems to eliminate the significant technical overhead associated with cryogenic liquid use.  These include a central cryogenic system that can supply cooling power to twelve independent locations to support LXe or LAr operation; a common slow control system; shareable xenon storage, circulation, and purification; radioactive gas source deployment (\emph{e.g.}, $^{85}$Kr, tritiated methane, radon) that can be injected into the noble gas stream; and support hardware, including an orbital tube welder up to 2” diameter, a small shop for fabrication, and a shared inventory of instrumentation, sensors, and vacuum and high-pressure hardware.  At present, the LNTF supports R\&D projects for the DUNE Near Detector, LZ upgrades (HydroX and Rn reduction via chromatography), and nEXO (Rn reduction via distillation).

Laboratory-scale facilities such as these serve a broad, unique and important role in service of the liquid noble detector development community, both lowering the cost of entry for developing new techniques and testing prototypes, and providing an excellent training ground for students and postdocs. In an era of increasingly large-scale experimental programs, these facilities and the efforts they support provide much needed opportunities for junior personnel to design and build whole experiments, while gaining valuable technical expertise from collaboration with the facilities' engineers and technical staff.









\bibliographystylened{JHEP}
\bibliographyned{Instrumentation/IF08/if08references} 







\end{document}